\documentclass[sigconf]{acmart}

\usepackage{tikz}
\newcommand{\ballnumber}[1]{\tikz[baseline=(myanchor.base)] \node[circle,fill=.,inner sep=1pt] (myanchor) {\color{-.}\bfseries\footnotesize #1};}

\usepackage[utf8]{inputenc}
\usepackage{xspace}
\usepackage{url}
\usepackage{booktabs}
\usepackage{graphicx}
\usepackage{multirow}
\usepackage{pgfplots}
\usepackage{pgfplotstable}
\usepackage{color, soul}
\newcommand{\ie}{\textit{i.e., \xspace}}
\newcommand{\eg}{\textit{e.g., \xspace}}

\newcommand{\toolname}{\textsc{AntiCopyPaster}\xspace}

\newcommand{\todo}[1]{\textcolor{red}{{\it [Old]}}}
\makeatletter
\newcommand\footnoteref[1]{\protected@xdef\@thefnmark{\ref{#1}}\@footnotemark}
\makeatother

\pgfplotstableread[col sep=comma,header=true]{% added header row
Category,1,2,3,4,5
User-Config,40,20,0,0,0
Clustering,40,20,0,0,0
Advanced,40,0,20,0,0
Multi-project,10,0,30,20,0
}\dataA

\pgfplotstablecreatecol[
 create col/expr={
    \thisrow{1} + \thisrow{2} + \thisrow{3} 
 }
]{sum}{\dataA}

\pgfplotsset{
  percentage plot/.style={
   point meta=explicit,
    every node near coord/.append style={
      font=\tiny,
      color=black,
    },
    nodes near coords={
      \pgfmathtruncatemacro\iszero{\originalvalue==0}% <-- needed to remove space
      \ifnum\iszero=0
      \pgfmathprintnumber[fixed,fixed zerofill,precision=1]{\pgfplotspointmeta}
      \fi
    },
    yticklabel=\pgfmathprintnumber{\tick}\,$\%$,
    ymin=0,
    ymax=100.01, % added .01 
    visualization depends on={y \as \originalvalue},
    enlarge x limits={abs=10mm}
  },
  percentage series/.style={
    table/x expr=\coordindex, %added
    table/y expr=(\thisrow{#1}/\thisrow{sum}*100),
    table/meta=#1
    }
}



\makeatletter
\newcommand{\linebreakand}{%
  \end{@IEEEauthorhalign}
  \hfill\mbox{}\par
  \mbox{}\hfill\begin{@IEEEauthorhalign}
}
\makeatother

\title[\toolname 2.0: Whitebox just-in-time code duplicates extraction]{\toolname 2.0: Whitebox just-in-time code duplicates extraction}
\author{Eman Abdullah AlOmar}
\affiliation{
    \institution{Stevens Institute of Technology}
    \city{Hoboken}
    \country{United States}
}
\email{ealomar@stevens.edu}

\author{Benjamin Knobloch}
\affiliation{
  \institution{Stevens Institute of Technology}
    \city{Hoboken}
    \country{United States}
}
\email{bknobloc@stevens.edu}

\author{Thomas Kain}
\affiliation{
    \institution{Stevens Institute of Technology}
    \city{Hoboken}
    \country{United States}
}
\email{tkain@stevens.edu}

\author{Christopher Kalish}
\affiliation{
    \institution{Stevens Institute of Technology}
    \city{Hoboken}
    \country{United States}
}
\email{ckalish@stevens.edu}

\author{Mohamed Wiem Mkaouer}
\affiliation{
    \institution{University of Michigan-Flint}
    \city{Michigan}
    \country{United States}
}
\email{mmkaouer@umich.edu}

\author{Ali Ouni}
\affiliation{
    \institution{ETS Montreal, University of Quebec}
    \city{Montreal, Quebec}
    \country{Canada}
}
\email{ali.ouni@etsmtl.ca}

\begin{document}

\begin{abstract}
\toolname is an IntelliJ IDEA plugin, implemented to detect and refactor duplicate code interactively as soon as a duplicate is introduced. The plugin only recommends the extraction of a duplicate when it is worth it. In contrast to current \textit{Extract Method} refactoring approaches, our tool seamlessly integrates with the developer's workflow and actively provides recommendations for refactorings. This work extends our tool to allow developers to customize the detection rules, \ie metrics, based on their needs and preferences. %\hl{Our experimental study ...}. 
 The plugin and its source code are publicly available on GitHub at
\url{https://github.com/refactorings/anti-copy-paster}. The demonstration video can be found on YouTube: \url{https://youtu.be/Y1sbfpds2Ms}.
%We developed a plugin for IntelliJ IDEA called \toolname, which tracks the pasting of code fragments inside the IDE and suggests the appropriate \textit{Extract Method} refactoring to combat the propagation of duplicates. Unlike the existing approaches, our tool is integrated with the developer's workflow, and pro-actively recommends refactorings. 
%Since not all code fragments need to be extracted, we develop a classification model to make this decision. When a developer copies and pastes a code fragment, the plugin searches for duplicates in the currently opened file, waits for a short period of time to allow the developer to edit the code, and finally inferences the refactoring decision based on a number of features.

%Our experimental study on a large dataset of 18,942 code fragments mined from 13 Apache projects shows that \toolname correctly recommends \textit{Extract Method} refactorings with an F-score of 0.82. Furthermore, our survey of 59 developers reflects their satisfaction with the developed plugin's operation.
%The plugin and its source code are publicly available on GitHub at 
%\url{https://github.com/JetBrains-Research/anti-copy-paster}. The demonstration video can be found on YouTube: \url{https://youtu.be/_wwHg-qFjJY}.

\end{abstract}
\maketitle
\section{Introduction}
%\hl{XXX}
\begin{comment}

\begin{figure*}
  \centering
  \includegraphics[width=.8\textwidth]{figures/pipeline.pdf}
  \vspace{-0.3cm}
  \caption{The pipeline of \toolname. \textit{Top}: training the model, \textit{bottom}: using the plugin.}
  \label{fig:pipeline}
  \vspace{-0.3cm}
 \end{figure*}

\end{comment}

\begin{figure*}
  \centering
  \includegraphics[width=.9\textwidth]{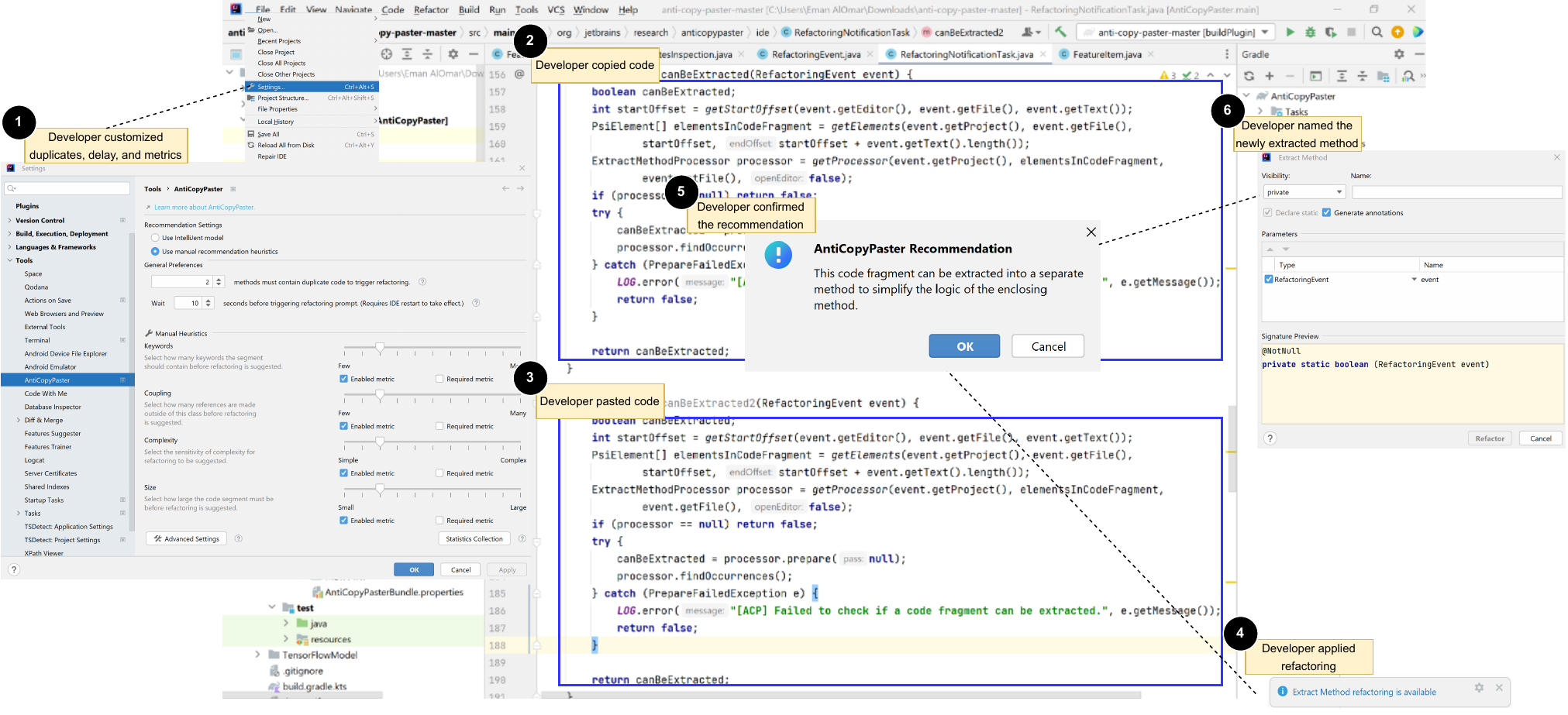}
  \caption{\textit{Extract Method} refactoring opportunity.}
  \label{fig:example}
  \vspace{-0.3cm}
 \end{figure*}

Duplicate code, also known as \textit{code clones}, is the act of creating copies of existing code fragments, with or without modification \cite{rattan2013software}. Several studies have shown how duplicate code hinders software maintenance, comprehension, and evolution \cite{thongtanunam2019will}. Additionally, duplicate code has become prevalent in GitHub, to the level of representing a significant challenge for all language models that are being trained using open source projects~\cite{lopes2017dejavu,allamanis2019adverse}. Therefore, refactoring duplicate code has emerged as a potential alternative to reduce its propagation~\cite{fanta1999removing}. Refactoring duplicate code consists of creating a method that contains one copy of the code fragment, and then replacing all its copies with a call to this newly created method. This refactoring is known as \textit{Extract Method} \cite{fowler2018refactoring,alomar2021preserving}. 

Although many studies proposed techniques to recommend method extraction for duplicate code~\cite{yoshida2019proactive,alcocer2020improving,hotta2010duplicate,alomar2023behind}, their adoption is constrained by being a posterior action that is only triggered upon the request of the developer and not when the duplicate code has been created. Also, these techniques need to exhaustively search the entire code base in search of refactorable clones, making the assumption that developers have the privilege to perform changes everywhere in the system. %to recommend proper \textit{Extract Method} refactorings. That is, the whole source code is used as input to list various \textit{Extract Method} refactorings for developers to apply. Such a solution makes a strong assumption that developers have the expertise of the entire system and the separate time to consider their options. 

In response to these limitations, we initially proposed \toolname \cite{alomar2022anticopypaster,alomar2023just}, an IntelliJ IDEA\footnote{IntelliJ IDEA: \url{https://www.jetbrains.com/idea/}} plugin that refactors duplicate code as soon as it is copy-pasted. The \textit{just-in-time} nature of the plugin allows the immediate recommendation of refactoring within the current context of the development, allowing the programmer to quickly convert the duplicated code into a method that is being called in all pasted locations in the code. The main challenge of our tool is that not all duplicate code needs refactoring; \ie the tool should recommend method extraction only when the refactoring is \textit{worth it}. To address this challenge, the decision of whether a duplicate code should be refactored was defined as a binary classification problem, where we trained a deep learning model that was trained on a large dataset of 18,942 previously performed \textit{Extract Method} refactorings. In a nutshell, for each duplicate code, we extract its corresponding syntactic and semantic model. This model is used to calculate a set of 78 comprehensive structural and semantic metrics, previously used in various studies recommending the \textit{Extract Method} refactoring ~\cite{shahidi2022automated,tiwari2022identifying,silva2015jextract,haas2016deriving,aniche2020effectiveness,van2021data,ouni2016multi}. The metrics' values are then fed into a convolutional neural network that learns existing patterns to distinguish between duplicate code fragments that are more likely and less likely to be extracted.

Although our proposed approach has shown promising results in terms of accuracy (F measure of 0.82), one of the main shortcomings reported by its users is the \textit{black-box} nature of the decision and the inability to change it, \ie developers cannot necessarily understand the rationale behind the decision and cannot change how it is being done either. Therefore, we extended our plugin to give developers full control of defining what the characteristics of duplicate code that need to be refactored are. To define the detection strategy of refactorable duplicates, the developer receives a comprehensive set of 78 metrics that cover the spectrum of the main structural categories, \ie `\textit{keywords}', `\textit{coupling}', `\textit{complexity}', and `\textit{size}'. Also, for the selected metrics, their threshold values can be easily indicated by providing the possibility of choosing any threshold between Min and Max, which are automatically calculated per metrics based on its statistical distribution in the entire project source set. These options elevate the limitation of existing tools that typically use 'fixed' threshold values, without a clear rationale for their division \cite{fontana2015automatic}. 

As an illustrative example, Figure \ref{fig:example}, shows a step-by-step scenario to automatically reproduce an \textit{Extract Method} refactoring. As mentioned in the change documentation, the intention of the refactoring is to reduce complexity. Therefore, we first open our plugin's settings \ballnumber{1}, and enable the complexity metric, along with setting its threshold to average, in order to flag any copy pasted code, whose cyclomatic complexity is higher than the average complexity of all methods in the system. Then, upon copy-pasting the code fragment \ballnumber{2}, \ballnumber{3}, the plugin calculates the metrics associated with the duplicate code, to identify that its complexity is higher than the accepted threshold, which triggers the refactoring recommendation through a pop-up notification \ballnumber{4}. If the recommendation is accepted \ballnumber{5}, the plugin invokes the IDE's \textit{Extract Method} refactoring with the appropriate parameters, and prompts the developer to input the name of the extracted method \ballnumber{6}.

%\hl{Now I need to see the updated approach to summarize it here} 

The preliminary evaluation of our tool shows its ability to reproduce more diverse instances of previously performed refactorings, which were collected using Refactoring Miner \cite{tsantalis2018accurate}, from open source projects.

\section{New Features}

\toolname aims to automatically provide \textit{just-in-time} \textit{Extract Method} refactoring recommendation as soon as duplicate code is created. Our tool takes various semantic and syntactic code metrics as input and makes a binary decision on whether the code fragment has to be extracted, either using our built-in binary classifier (previous version \cite{alomar2022anticopypaster,alomar2023just}), or using user-specified configuration (this extension). In our revision of \toolname, we sought to significantly enhance the usability and versatility of the plugin. Listed below are the objectives we pursued in our work on \toolname:
\vspace{-.3cm}
\begin{itemize}
    \item Users will be able to set individual metrics as enabled and/or required in the settings menu.
\item \toolname will be able to measure and compare 4 metrics: `\textit{keywords}', `\textit{coupling}', `\textit{complexity}', and `\textit{size}',
\item Users will be able to customize how many instances of duplicate code are required to trigger a notification in the settings menu., as well as how long the buffer time between a paste action and its notification event is.
\item Users will be able to open an advanced settings menu from the settings menu, where each metric can be customized.
For `\textit{keywords}', users will be able to choose between total keyword count and keyword density, as well as which specific keywords are enabled. For `\textit{coupling}', users can choose between total coupling and coupling density, as well as choose between total connectivity, field connectivity, and method connectivity.
For `\textit{complexity}', users will be able to choose between the total area of the segment, area density, method declaration area, and method declaration depth density. For `\textit{size}', users will be able to choose between number of lines, number of symbols, and density of symbols, as well as between the size of the segment and the size of the method declaration.
\item \toolname shall determine if a notification event is triggered by comparing the stored group of metrics with thresholds determined by sensitivity percentile calculations.
\item \toolname shall run separately across all IntelliJ project windows that are open at once, including allowing different settings in each window and triggering separate notifications.
\end{itemize}

\begin{figure}
  \centering
  \includegraphics[width=.9\columnwidth]{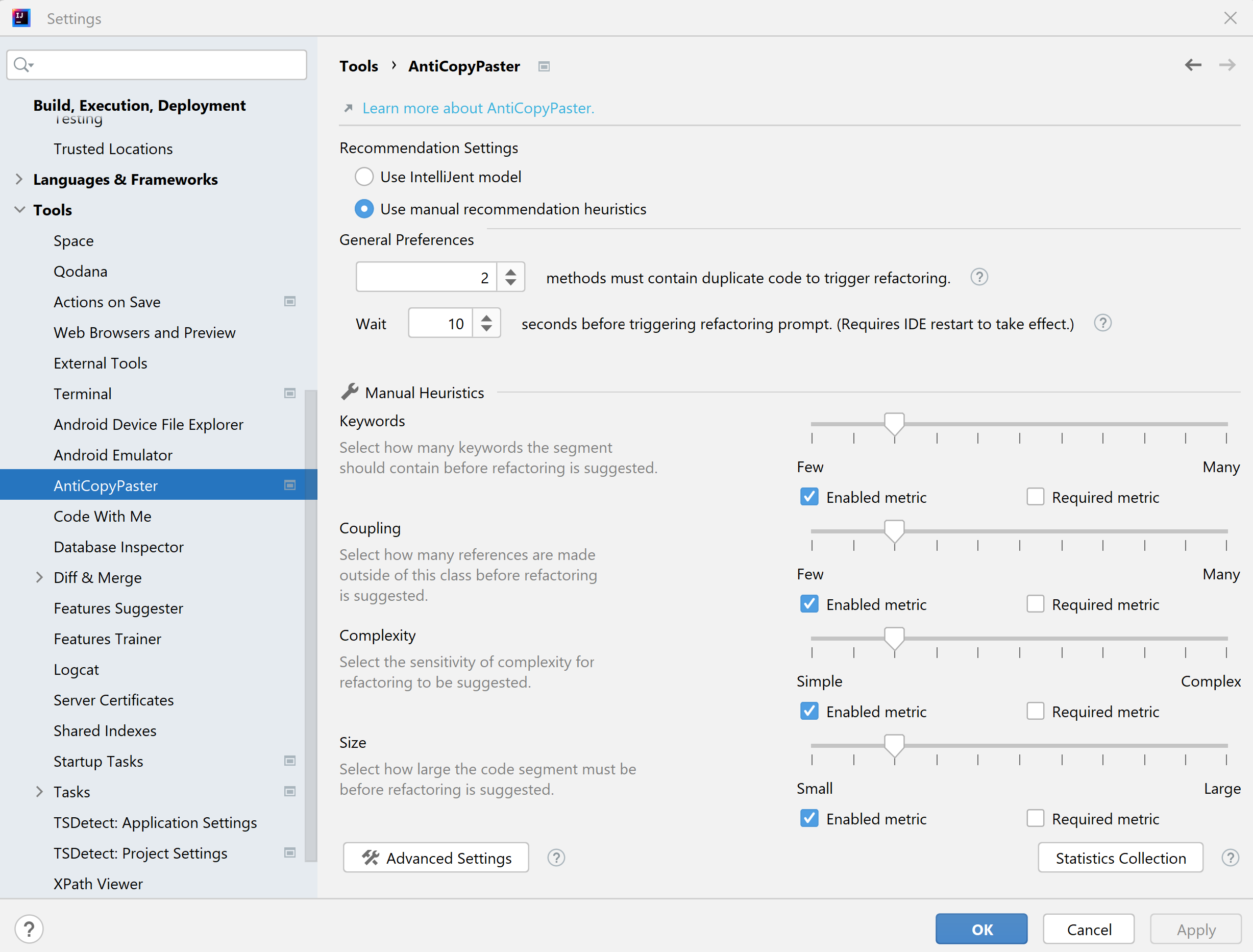}
  \caption{ \toolname metrics clustering feature.}
  \label{fig:clustering}
  \vspace{-0.3cm}
 \end{figure}

 \begin{figure}
  \centering
  \includegraphics[width=.9\columnwidth]{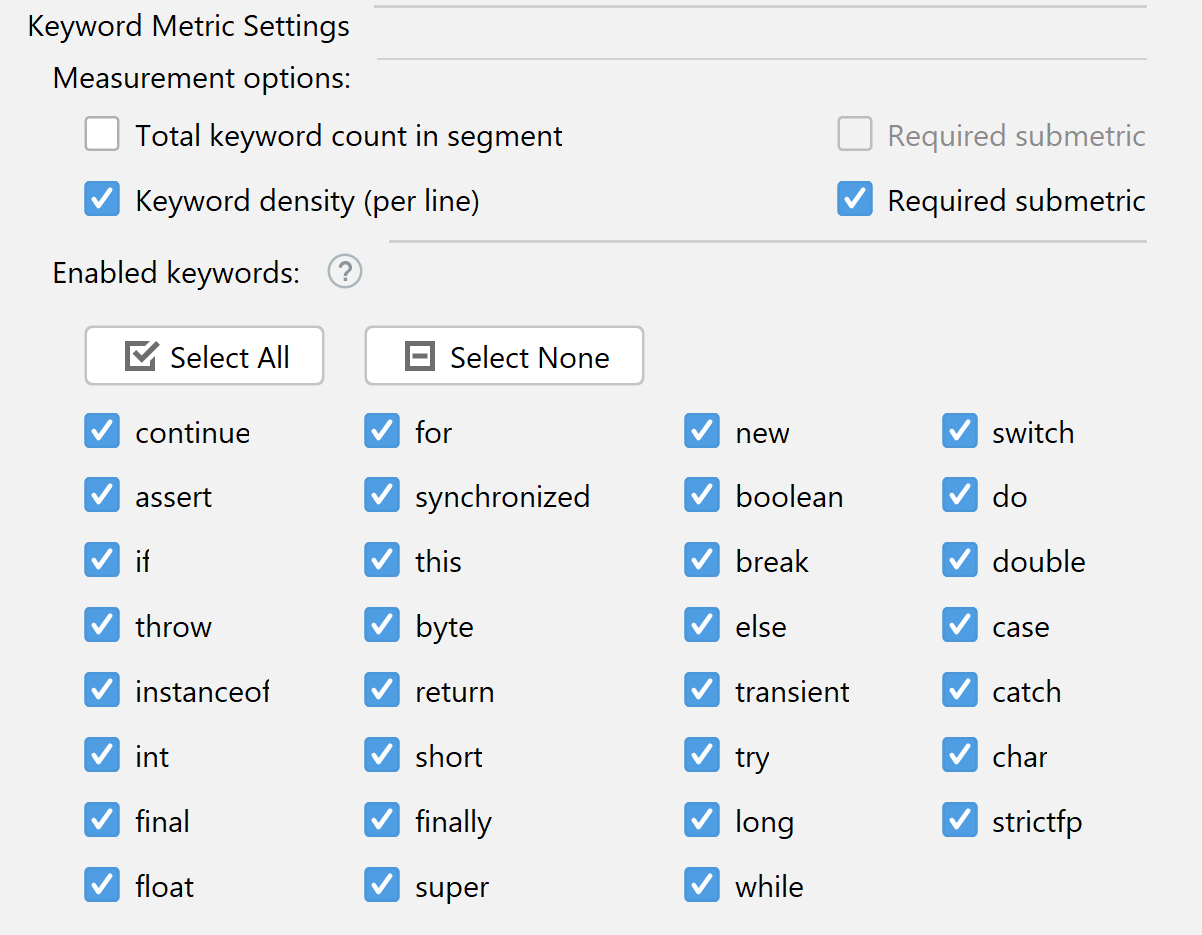}
  \caption{ \toolname metrics advanced feature (Keyword).}
  \label{fig:keyword}
  \vspace{-0.3cm}
 \end{figure}

 \begin{figure}
  \centering
  \includegraphics[width=.9\columnwidth]{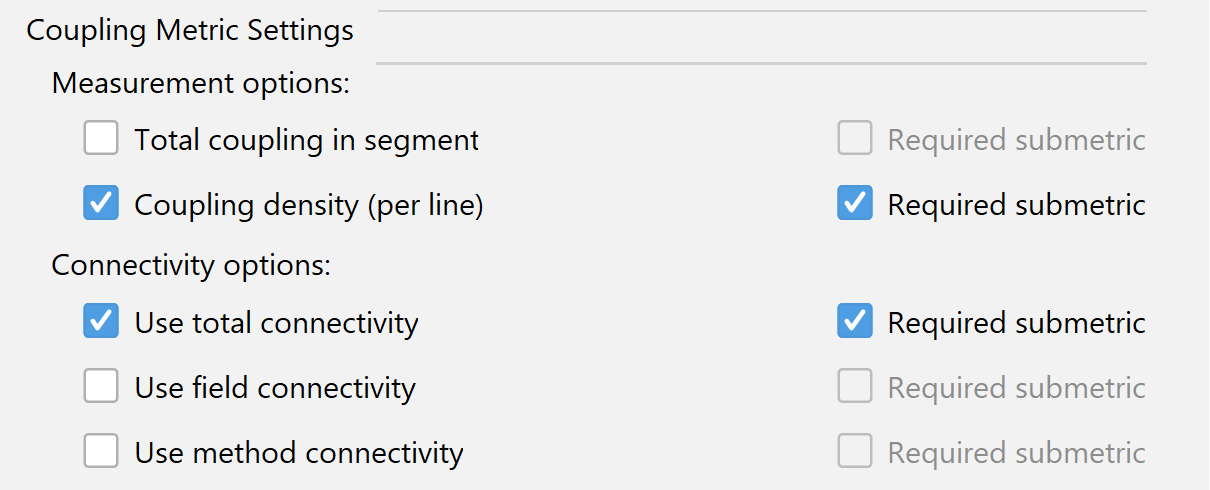}
  \caption{ \toolname metrics advanced feature (Coupling).}
  \label{fig:coupling}
  \vspace{-0.3cm}
 \end{figure}

 \begin{figure}
  \centering
  \includegraphics[width=.9\columnwidth]{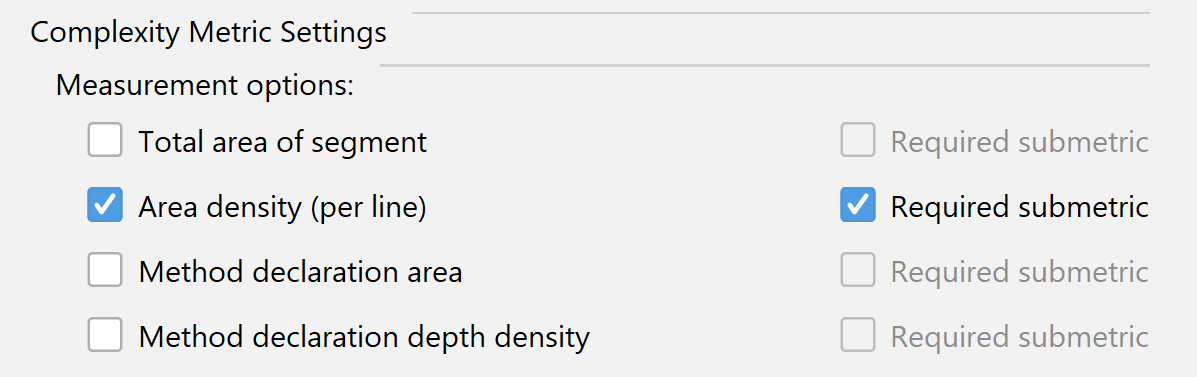}
  \caption{ \toolname metrics advanced feature (Complexity).}
  \label{fig:complexity}
  \vspace{-0.3cm}
 \end{figure}

 \begin{figure}
  \centering
  \includegraphics[width=.9\columnwidth]{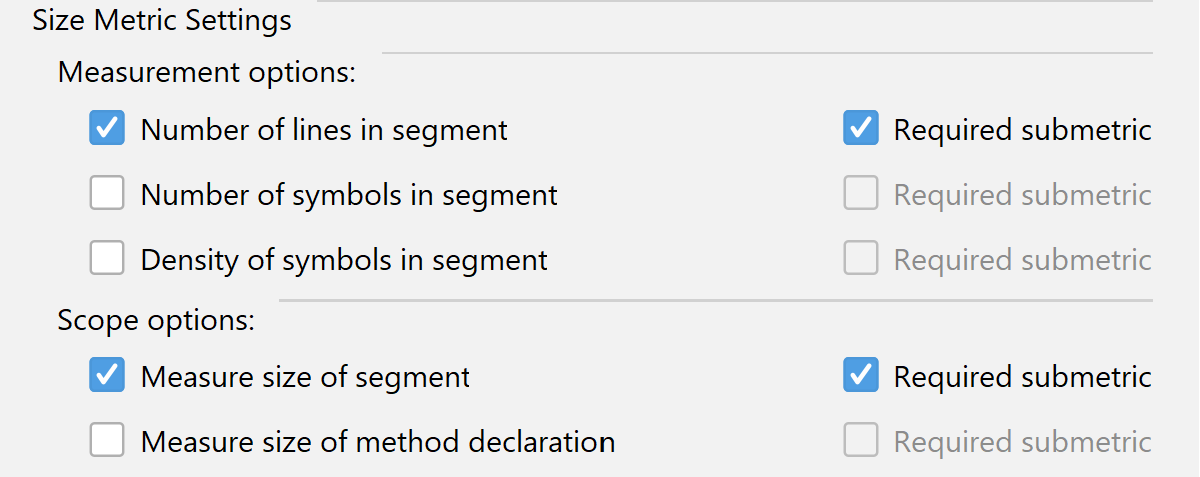}
  \caption{ \toolname metrics advanced feature (Size).}
  \label{fig:size}
  \vspace{-0.3cm}
 \end{figure}

Next, we elaborate on these features that enhance the user experience with \toolname.

%\textbf{Code Metrics Selection}. 

 \subsection{Percentile-Based Sensitivity Values}
  One of the significant components of the extended feature was the implementation of the plugin’s percentile-based system following the introduction of sliders to denote metric sensitivities.  When our interface changes included allowing for sensitivity values along a slider (\ie a numerical scale), we allowed a number between 1 and 100 corresponding to a percentile value.  

 \subsection{Metrics Clustering}

The identification of refactoring opportunity is metric-based; \eg if a developer's preference is to prevent the propagation of long duplicates, they can specify size related metrics to be used (such as lines of code, number of literals, etc.). The selected metrics, along with their corresponding threshold, constitute the detection rule to identify the refactoring opportunity. The plugin offers the selection among 78 metrics that have been extensively used in previous studies~\cite{aniche2020effectiveness,haas2016deriving}.
 %In the original version of the plugin, the model would automatically gather values for 78 different metrics that form the basis for the refactoring suggestions. 
 A key objective of the extension was to encompass all metrics while allowing users the flexibility to customize the metrics they preferred in the cluster. An overview of the clustering interface is depicted in Figure \ref{fig:clustering}.

 \subsection{Duplicate and Delay Setting}
In addition, above the metric sliders we added numerical text entry boxes for users to denote how many methods must contain code duplicates for a notification to trigger and how long the delay between paste events and their corresponding notifications should be. In the original version, there must be code duplicates in two separate methods to trigger a notification, and there would be a delay of 10 seconds after code was pasted before a notification was triggered. %As with the metric sliders and checkboxes, we maximize customization to enhance the user experience with \toolname.

 \subsection{Advanced Settings}

 An advanced settings page was established on the plugin settings interface for users to customize how each metric should be interpreted by the plugin. The advanced settings menu can be opened from a button on the regular settings menu, which opens a new window with the advanced settings on it.

 The `\textit{keywords}' metric (see Figure \ref{fig:keyword}) has two categories for customization within the advanced settings menu. First, the measurement options allow the user to select if the plugin should measure the total keyword count or the keyword density in a segment of code. Both choices may be selected, and in addition, both have the option to be set as a required submetric. The second prompt for the keyword metric is the enabled keywords section, which lists 31 unique Java keywords. By default, all 31 are selected, though the user has the option to allow only a select subset to be recognized by the plugin. For convenience, there are buttons to select all and select none of the keywords as well. This setting also features a help icon next to it that provides clarification on enabling keywords.

 The `\textit{coupling}' metric (see Figure \ref{fig:coupling}) has two categories for customization within the advanced settings menu. First, the measurement options provide the same choice as the measurement options category in the keyword metric settings. The second prompt for the coupling metric is the connectivity options section, which lets the user choose whether the plugin should use total, field, or method connectivity. As with the measurement options, any number of these may be selected at once, and all have the option to be required.

 The `\textit{complexity}' metric (see Figure \ref{fig:complexity}) has only one category for customization within the advanced settings menu, that being the measurement options. Users may select whether the plugin should use the total area, area density, method declaration area, or method declaration depth density of a segment of code to determine complexity. Any number of these may be selected at once, and all have the option to be required.

 The `\textit{size}' metric (see Figure \ref{fig:size}) has two categories for customization within the advanced settings menu. First, the measurement options, which let the user select whether the plugin should use the number of lines, the number of symbols, or the density of symbols in a segment of code to determine its size. The second prompt for the size metric is the scope options section, which lets the user choose whether the plugin should measure the size of code segments or method declarations. For both of these categories, any number of choices may be selected at once and all have the option to be required.

Moreover, we added two features to clarify how the settings work for new users: a link at the top of the menu that directs to the official website of the plugin where the metric settings are explained, and help icons next to the two numerical text entries that further explain those customization settings.

 %\vspace{0.2cm}
\vspace{-.3cm}
  \subsection{Multi-Project Support}
 The original version of \toolname does not function on multiple open projects simultaneously. Many classes within its codebase did not account for the assumption that they may be needed on a per-project basis. Some systems, such as the plugin’s notification system, only acted relative to the first open project in the IDE. In the expanded feature, we realized that adapting the plugin to work with multiple projects would be necessary, as this behavior was otherwise undocumented and unintuitive.

\section{Pleminary Evaluation}

\textbf{Correctness.} To evaluate the correctness of our extension, we selected a dataset of already refactored clones, in order to see if we can reproduce their refactoring, via our plugin. To do so, we started with a dataset of refactorings \cite{msr2022}, detected by  Refactoring Miner \cite{tsantalis2018accurate}. Since these code changes can be driven by various intents, we filtered them by keeping only method extractions whose commit messages explicitly mention the removal of duplicate code or clones. Also, for each selected refactoring, we need to checkout the project, and manually set all its dependencies, since the code needs to compile in order for the plugin to function. We ended up using 10 refactoring instances for our preliminary sanity check. When testing every instance with both versions of the plugin, we were able to reproduce 7 instances by selecting the appropriate thresholds from the settings menu (second version), while the model, implemented in the first version, was able to flag 5 instances. The ability to reproduce more cases by the second version shows the value of adding self-configuration that allows one to better mimic human decisions, which may or may not follow the \textit{wisdom of the crowd} (used to train the model of the first version). As for the cases that we missed, they are of particular interest to us because they can help us improve our tool in the future. The 3 instances missed by both versions are because the duplicate code was not exact, whereas our tool only works with type-1 clones.

\begin{figure}
  \centering
  \includegraphics[width=.9\columnwidth]{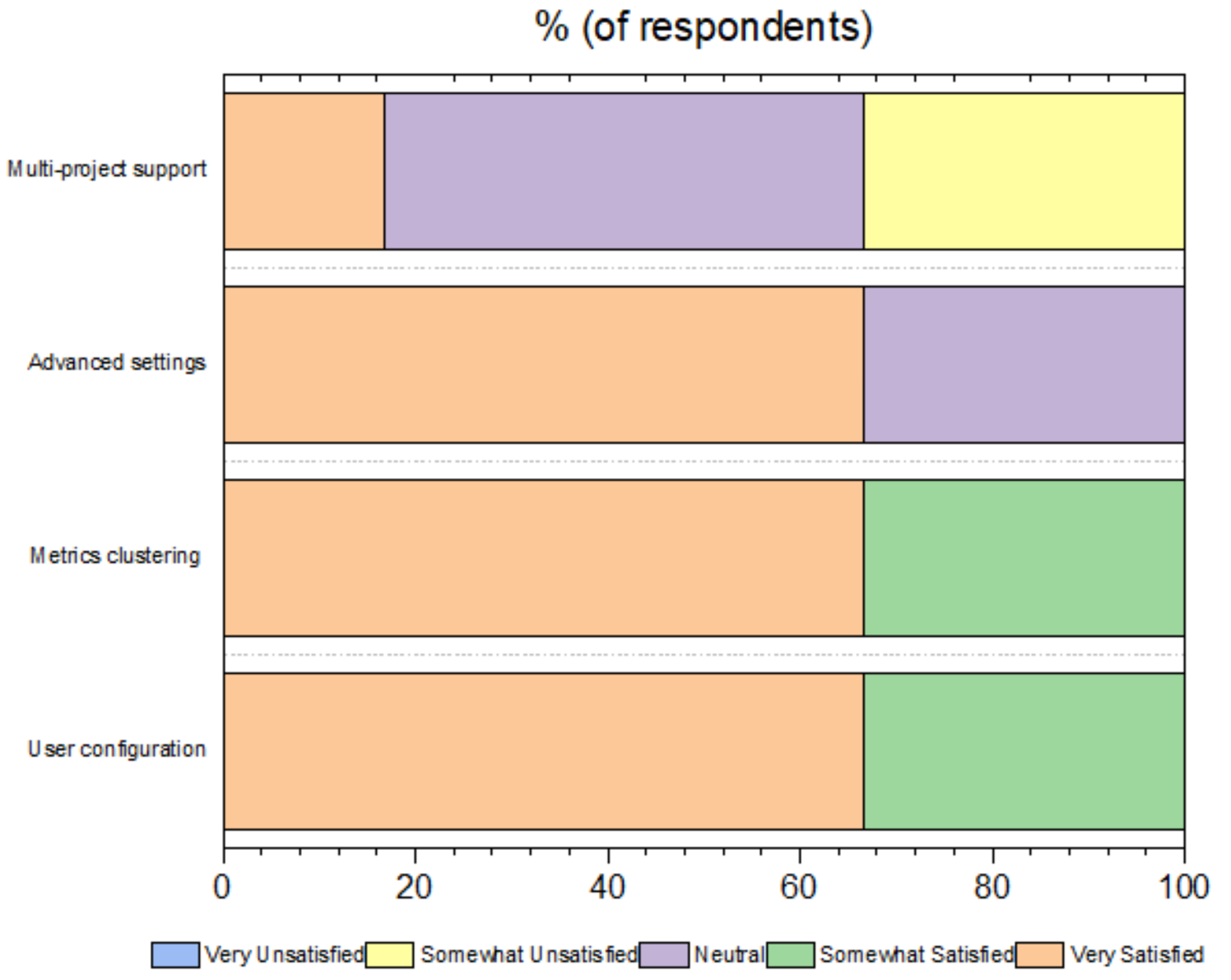}
\caption{Participants’ satisfaction with various aspects of
the \toolname tool}
\label{fig:usefulness}
  \vspace{-0.3cm}
 \end{figure}

\textbf{Usefulness.} With the release of the second version, we posted a survey for users to optionally take it. We report the results of the users who have taken it in Figure \ref{fig:usefulness}. The survey results varied from 'Very Satisfied', mainly with the new feature of user-configurability, and lesser with how metrics were outlined in the settings. On the other hand, users are less satisfied with the advanced menu setting and even unsatisfied with the multi-project copy-pasting code support.

\bibliographystyle{ACM-Reference-Format}
\balance
\bibliography{cites}

\end{document}